\documentclass[letterpaper]{aa}
\usepackage{graphicx}
\usepackage{natbib}
\usepackage{txfonts}
\begin{document}
   \headnote{Research Note}
   \title{Astrometric proof of companionship \\ for the L dwarf companion candidate GJ 1048B}
   \subtitle{}

   \author{Andreas Seifahrt \and Ralph Neuh\"auser \and Markus Mugrauer}
          
   \institute{Astrophysikalisches Institut, Universit\"at Jena, Schillerg\"asschen 2-3, 07745 Jena, Germany}
   \offprints{Andreas Seifahrt, \email{p7sean@astro.uni-jena.de}}
   \date{Received February 2004 / Accepted 14 March 2004}

   \abstract{\citet{gizis} reported a companion candidate of spectral type L1 near the K2 dwarf GJ 1048 using the Two Micron All-Sky Survey (2MASS). At that time it was not possible to verify companionship astrometrically using the 2MASS data alone due to the small proper motion of GJ 1048. We now show that both objects share the same proper motion by using data from the UK Schmidt Telescope Near-infrared (IVN) Southern Survey as the first epoch and data from 2MASS as the second epoch. Our technique of subtracting the PSF of the primary from the SuperCOSMOS I scans of the Southern Survey enables the astrometry of the companion candidate to be measured directly.

   }

   \maketitle
%

\section{Introduction}

Brown dwarfs are objects with masses below the mass limit for core hydrogen burning. They encompass a range of spectral types, from late M, for young brown dwarfs, down to the new spectral types L and T, defined by \citet{kirk99}, \citet{martin99} and \citet{burg00}. So far a considerable number of such objects has been found, mainly by infrared (IR) surveys like the Two Micron All-Sky Survey (hereafter 2MASS) and the DEep Near Infrared Survey (hereafter DENIS). Although many field brown dwarfs are known, few are known as companions to stars. The majority of these stars are objects in the solar neighborhood and show a high proper motion. Confirmation of companionship can therefore be realized by determining a common proper motion of the star and the companion candidate. The significance and ease of this method generally increases with increasing proper motion and epoch difference between the observations. 

~\citet{gizis} identified the 2MASS object 2MASSI J0235599--233120 as a companion candidate 11\farcs9 separated from the K2 dwarf GJ1048. The JHK magnitudes of the object were consistent with an early L dwarf at the same (HIPPARCOS measured) distance of $21.3\pm0.47$~pc as GJ1048. Thus, it is likely that the object is a companion of GJ1048. Gizis et al. referred to the companion as GJ1048B, the primary as GJ1048A. If GJ1048B is indeed a companion, then its mass (at the same age and distance as the primary) is near the sub-stellar limit \citep{gizis}. 

Due to the comparably small proper motion of GJ1048 ($\mu_{\alpha}=84.53\pm0.98~$mas/yr, $\mu_{\delta}=13.04\pm0.76~$mas/yr) it has so far not been possible to verify the companionship astrometrically. Since the epoch difference between the two major IR surveys 2MASS and DENIS is only in the range of a few years, stars showing only moderate proper motion like GJ 1048 are not suitable for a astrometric comparison within both mentioned surveys. It is therefore helpful to look for older red or IR surveys, sensitive enough to allow the detection of the companion candidates. The strength of this method is demonstrated by the discovery of $\epsilon$ Indi B \citep{scholz} on SuperCOSMOS Sky Survey images. 

We report the missing confirmation of common proper motion companionship for GJ1048B. 

\section{GJ1048A\&B on the SuperCOSMOS Sky Survey}

Several thousand photographic plates from Schmidt surveys were scanned by SuperCOSMOS. The SuperCOSMOS Sky Survey (hereafter SSS) contains data from the UK Schmidt Telescope Southern Survey plates. The sensitivity of these plates is 19.5 mag in I-band. The pixel size is 0\farcs7 (10 micron).

Sensitivity and resolution are high enough to detect GJ1048B but the high contrast ratio between GJ1048A and B leads to serious light contamination by the PSF of GJ1048A. To allow a better position fit, special image processing techniques are applied.

\begin{figure}
   \resizebox{\hsize}{!}{\includegraphics[bb=0 0 697 578,clip]{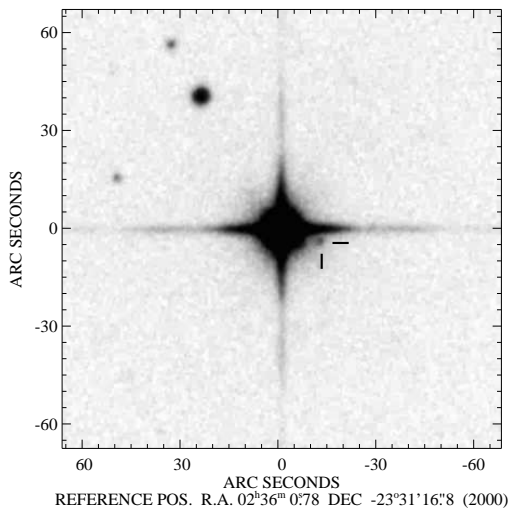}}
   \caption{SSS image of GJ1048. GJ1048B (marked) lies close to the diffraction spike of the primary. North is up, east to the left.}
   \label{GJ1048A}%
    \end{figure}

Fig.~\ref{GJ1048A} shows the companion candidate within the PSF of GJ1048A. A detection of GJ1048B via GAIA and SExtractor \citep{sextractor} is not possible. The object is also not contained in the SSS NIR catalog. To allow a detection of the companion candidate a method has to be used to suppress the PSF of the primary. 

A possible technique uses the rotational symmetry of the star profile. An image stack is built of images rotated around the center of the primary and subtracted from the original image. The median of all these images is nearly free of the primary. 

\begin{figure}
   \resizebox{\hsize}{!}{\includegraphics[bb=0 0 697 578,clip]{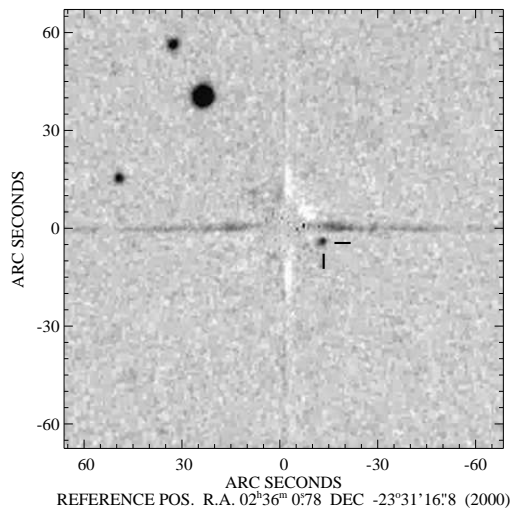}}
   \caption{SSS image of GJ1048 after PSF subtraction. GJ1048B (marked) is now easily detectable. North is up, east to the left.}
   \label{GJ1048B}%
    \end{figure}

For GJ1048 this technique has to be adapted to suppress the diffraction spikes near the object. Rotation angles in 1$\degr$ steps in 5$\degr$ intervals around 90$\degr$, 180$\degr$, 270$\degr$ and 360$\degr$ show the best results. The signal-to-noise ratio increases from $\sim3$ to $\sim6$ for GJ1048B allowing automatic source detection. The suppression of the diffraction spikes allows a more precise position fit. Due to the non-uniformity of the diffraction spikes the success of this technique is limited but satisfying (see Fig.~\ref{GJ1048B}). The center of rotation has generally a strong influence on the shape of the profile of the companion and thus on its position. However position errors resulting from this influence do not contribute much to the overall position fit error of the companion candidate in this case. The major direction of proper motion is in Right Ascension. The error due to the mentioned above effects is only notable in Declination, because of the strong influence of the nearby diffraction spike. Therefore error propagation shows only a minor increase of the overall error in both positions.

\section{Data Analysis: Astrometry}

In a further step, an astrometric fit of the 2MASS catalog is applied to the SSS image to use the 2MASS catalog as a reference for comparing astrometry. 
This is done using the astrometric calibration routine of GAIA. The result is a good coordinate concordance within both images with remaining errors of approximately 30~mas in Right Ascension ($\alpha$) and 130~mas in Declination ($\delta$). The position fit of all objects is realized via SExtractor resulting in a list of extracted star positions, including fit errors.

At this stage several methods can be used to show the common proper motion of the two objects. 

One possibility is to directly measure the position differences in $\alpha$ and $\delta$ for all stars in the image. Co-moving objects should have consistent position differences in $\alpha$ and $\delta$ (overlapping error ellipses) and should be well separated from the field stars, which ideally should show no significant motion at all. This method needs no assumptions on the nature of the primary or the secondary but suffers from comparatively large errors due to image distortions either from the instrument itself or from the plate scan. These errors increase with larger images.

\begin{figure}
   \resizebox{\hsize}{!}{\includegraphics[bb=14 14 269 195,clip]{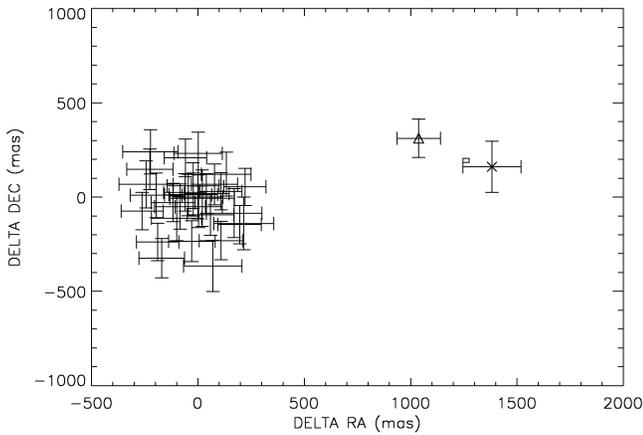}}
   \caption{Measured proper motion between SSS image (1983/12/12) and 2MASS catalog (1998/11/11). Proper motion derived from HIPPARCOS data shown as small box near GJ1048B, GJ1048A marked as triangle and GJ1048B as cross. 1$\sigma$ errors plotted for all values.}
   \label{pm}%
\end{figure}

For GJ1048 the results show that the companion candidate shares the HIPPARCOS measured proper motion of the primary within the 1~$\sigma$ errors, while GJ1048A is consistent with this value by $\sim2.5~\sigma$ (see Fig.~\ref{pm}). 

The proper motion distribution of the field stars lie within a halo of $\sim3~\sigma$ around zero. While the last value reflects the effect of the discussed image distortion and is in this sense a limit for the reliability of this measurement, the comparatively high error for GJ1048A is caused by higher fit errors due to the saturation of the primary. Both GJ1048A and GJ1048B are by $\sim10~\sigma$ different from stationary (zero motion) background objects.
\begin{figure}
   \resizebox{\hsize}{!}{\includegraphics[bb=14 14 260 191,clip]{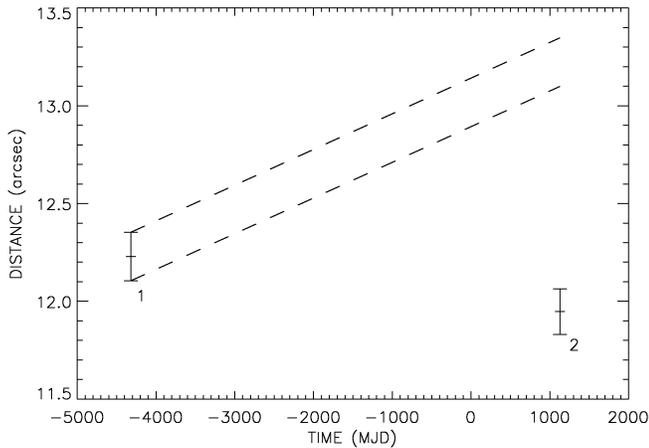}}
   \caption{Measured separation between GJ1048A and GJ1048B in (1) SSS image (1983/12/12) and (2) 2MASS catalog (1998/11/11). The dashed line indicates the separation that would have been measured if GJ1048A has the HIPPARCOS measured proper motion and GJ1048B is a stationary background object. 1~$\sigma$ errors are shown.}
   \label{distance}%
\end{figure}
The second possible method is to measure the separation between the primary and the secondary in both images and compare this value with a theoretical value derived under the assumption that the companion candidate is a stationary background object. Because this method does not make any use of the field stars, only small image portions are measured and the error due to large-scale image distortion does not effect the results. The resulting errors are therefore smaller and the result is more significant. However it should kept in mind that for systems that include a notable angle between the direction of proper motion and the line connecting the objects some information loss occurs in measuring only the absolute separation rather than the position difference in two coordinates ($\alpha$ and $\delta$). To compensate for the loss in significance of these measures, both the separation and the position angle has to be measured.
\begin{table}[b!]
\caption[DATA]{Data for GJ1048. Errors for distance and position angle are 1~$\sigma$ errors.}
\label{values}
\vspace{.0 cm}
\resizebox{!}{!}{
\begin{centering}
\begin{tabular}{cccccccc}
\hline
\hline Epoch & \textbf{Source of} & \multicolumn{2}{c}{GJ1048A} & \multicolumn{2}{c}{GJ1048B}& Separation & Pos.Angle \\ 
& \textbf{Astrometry} & R.A. & Dec & R.A. & Dec & & \\
\hline
1983.949 & \textbf{SSS} & 02$^{\rm h}$ 36$^{\rm m}$ 00\fs69 &$-23$\degr 31\arcmin 17\farcs33 & 02$^{\rm h}$ 35$^{\rm m}$ 59\fs85 &$-23$\degr 31\arcmin 20\farcs96 & 12\farcs23 $\pm$ 0\farcs12 & 252\fdg72 $\pm$ 0\fdg58 \\ 
1998.864 & \textbf{2MASS} & 02$^{\rm h}$ 36$^{\rm m}$ 00\fs76 &$-23$\degr 31\arcmin 16\farcs74 & 02$^{\rm h}$ 35$^{\rm m}$ 59\fs93 &$-23$\degr 31\arcmin 20\farcs52 & 11\farcs95 $\pm$ 0\farcs12 & 251\fdg54 $\pm$ 0\fdg56   \\ 
\hline 
\end{tabular} \end{centering}}
\end{table} 
For GJ1048 the separation between both objects stays constant within less than 2~$\sigma$. The difference between the measured distance and the calculated value (13\farcs22 $\pm$ 0\farcs12) for a stationary background object is notable. 
The background hypothesis can therefore be rejected by more than 7 $\sigma$ (see Fig.~\ref{distance}).
Due to the fact that the position angle of the object is nearly the same for the direction of proper motion, the error in position angle (see Tab.~\ref{values}) cannot be taken into account.

The effect of parallax is some 0\farcs022 for the given epoch difference and can be easily neglected. Orbital motion is however not negligible. Assuming the extreme case of a face-on orbit, a physical separation of 250 AU and a mass of 0.75 M$_{\sun}$ for GJ1048A, a basic estimate of the maximum orbital motion in the epoch difference of nearly 15 years is 1\fdg2 causing a motion on the sky of 0\farcs25 perpendicular to the direction of proper motion but only 0\farcs0025 in the direction of proper motion. This maximum orbital motion is of the order of the derived 1~$\sigma$ error of the position angle, and for the separation it is well below this margin. To verify orbital motion at least another 30 years have to pass before a change of the position angle by more than 3~$\sigma$ can be observed.

\section{Conclusions}
Both methods presented in this paper show a consistent result for the common proper motion of GJ1048 A and B.  In combination with the previous results from \citet{gizis} there can be no reasonable doubt that GJ1048B is physically associated with GJ1048A. Therefore the probability that GJ1048B is not a companion of GJ1048A is very small. However the final proof of this result has to be postponed until more of the orbital motion is detectable.

\begin{acknowledgements}
This publication makes use of data products from the Two Micron All Sky Survey (which is a joint project of the University of Massachusetts and the Infrared Processing and Analysis Center/California Institute of Technology, funded by the National Aeronautics and Space Administration and the National Science Foundation) and data based on the SuperCosmos Sky Surveys at the Wide Field Astronomy Unit of the Institute of Astronomy, University of Edinburgh. We have also used the VizieR catalog access tool, CDS, Strasbourg. 
\end{acknowledgements}

\end{document}